\documentclass{article}

\usepackage{PRIMEarxiv}

\usepackage[utf8]{inputenc} % allow utf-8 input
\usepackage[T1]{fontenc}    % use 8-bit T1 fonts
\usepackage{hyperref}       % hyperlinks
\usepackage{url}            % simple URL typesetting
\usepackage{booktabs}       % professional-quality tables
\usepackage{amsfonts}       % blackboard math symbols
\usepackage{nicefrac}       % compact symbols for 1/2, etc.
\usepackage{microtype}      % microtypography
\usepackage{lipsum}
\usepackage{graphicx}
\graphicspath{{media/}}     % organize your images and other figures under media/ folder

\usepackage{float}

\usepackage{tabularx}
\usepackage[table]{xcolor}
\usepackage{geometry}
\usepackage{listings}
\usepackage{xcolor} % for colors (optional but useful)

\lstdefinelanguage{SQL}{
  morekeywords={SELECT, FROM, WHERE, GROUP, BY, ORDER, AS, AVG, TO_CHAR},
  sensitive=false,
  morecomment=[l]{--},
  morestring=[b]',
}

\title{Accelerating Drug Discovery Through Agentic AI: A Multi-Agent Approach to Laboratory Automation in the DMTA Cycle
}

\author{
\parbox{\textwidth}{
\centering
Yao Fehlis, Charles Crain, Aidan Jensen, Michael Watson, James Juhasz, Paul Mandel, Betty Liu, Shawn Mahon, Daren Wilson, Nick Lynch-Jonely, Ben Leedom, David Fuller\\
Artificial, Inc.
}
}

\begin{document}
\maketitle

\begin{abstract}
The pharmaceutical industry faces unprecedented challenges in drug discovery, with traditional approaches struggling to meet the demands of modern therapeutic development. This paper introduces Tippy, a novel agentic AI framework that transforms laboratory automation through specialized AI agents operating within the Design-Make-Test-Analyze (DMTA) cycle. Our multi-agent system employs five specialized agents—Supervisor, Molecule, Lab, Analysis, and Report, with Safety Guardrail oversight—each designed to excel in specific phases of the drug discovery pipeline. As far as we know, Tippy represents the first production-ready implementation of specialized AI agents designed specifically for automating the DMTA cycle, providing a concrete example of how agentic AI can transform laboratory workflows. By leveraging autonomous AI agents that can reason, plan, and collaborate, we demonstrate how Tippy's agentic AI can accelerate DMTA cycles while maintaining the scientific rigor essential for pharmaceutical research. The system shows significant improvements in workflow efficiency, decision-making speed, and cross-disciplinary coordination, offering a new paradigm for AI-assisted drug discovery.
\end{abstract}

% keywords can be removed

\section{Introduction}
\subsection{The Challenge of Modern Drug Discovery}
The pharmaceutical industry stands at a critical juncture. Despite advances in computational chemistry, high-throughput screening, and molecular biology, the cost and time required to bring new drugs to market continue to escalate. The traditional drug discovery process, often taking 10-15 years and costing billions of dollars, struggles with inefficiencies rooted in siloed workflows, manual data integration, and sequential rather than parallel decision-making\cite{zhang2025artificial,hopkins2025academic,joshi2025harnessing}.

At the heart of modern pharmaceutical research lies the Design-Make-Test-Analyze (DMTA) cycle—an iterative framework that drives the optimization of drug candidates from initial concept to clinical development. The DMTA cycle represents both the fundamental methodology of medicinal chemistry and the primary bottleneck in drug discovery efficiency. Each cycle involves designing new molecular entities, synthesizing them in the laboratory, testing their properties through various analytical methods, and analyzing the results to inform subsequent design decisions.

\subsection{Understanding the DMTA Framework}
The Design-Make-Test-Analyze cycle\cite{wesolowski2016strategies,medcalf2024overcoming,ghiandoni2024augmenting} represents the core methodology of modern medicinal chemistry, providing a structured approach to drug optimization that has guided pharmaceutical research for decades. Each phase of the DMTA cycle addresses specific aspects of drug development, from molecular design through experimental validation to data interpretation and hypothesis refinement.

The Design phase encompasses computational chemistry, molecular modeling, and structure-based drug design. Medicinal chemists leverage tools ranging from quantum chemistry calculations to machine learning models to propose new molecular entities with improved properties\cite{cavasotto2019computational,hu2023deep,zhu2025identifying,young2009computational,xu2024current,leelananda2016computational}. This phase requires deep understanding of structure-activity relationships (SAR), pharmacokinetics, and target biology to generate hypotheses that can be tested experimentally.

The Make phase involves synthetic chemistry, process optimization, and compound preparation. Synthetic chemists develop efficient routes to target molecules, optimize reaction conditions, and scale synthesis as needed. This phase requires expertise in organic chemistry, reaction mechanisms, and analytical characterization to ensure compound identity, purity, and consistency.

The Test phase encompasses a broad range of analytical and biological assays designed to characterize compound properties. From basic physicochemical measurements like solubility and LogP to complex cell-based assays and animal studies, this phase generates the experimental data needed to evaluate compound performance against target criteria.

The Analyze phase involves data interpretation, pattern recognition, and decision-making about next steps. Researchers must integrate results from multiple assays, identify trends and outliers, and formulate hypotheses for the next design iteration. This phase requires statistical analysis, data visualization, and scientific reasoning to extract actionable insights from complex datasets.

\subsection{Current Limitations in DMTA Implementation}
Despite its conceptual elegance, the DMTA cycle faces significant implementation challenges that limit its effectiveness in modern drug discovery. Traditional approaches suffer from several critical bottlenecks that impede rapid iteration and efficient resource utilization.

Sequential rather than parallel execution represents a major inefficiency in current DMTA implementations. Most organizations execute DMTA phases in series, waiting for complete results from one phase before initiating the next. This approach, while scientifically conservative, creates significant delays and underutilizes available resources. Modern drug discovery demands more parallel and overlapping execution to maximize throughput and minimize cycle times.

Data integration and communication barriers pose another significant challenge. Each phase of the DMTA cycle generates distinct types of data using different tools, formats, and analytical frameworks. Medicinal chemists work with molecular structures and SAR data, synthetic chemists focus on reaction yields and analytical characterization, and biologists analyze dose-response curves and mechanism studies. Integrating these diverse data types into coherent decision-making frameworks remains a manual and error-prone process.

Resource coordination and scheduling inefficiencies\cite{fehlis2025uncoveringbottlenecksoptimizingscientific} further compound these challenges. Modern pharmaceutical research involves multiple specialized teams, expensive instrumentation, and complex experimental protocols. Coordinating these resources across DMTA cycles requires significant project management overhead and often results in suboptimal utilization of both human expertise and analytical capabilities.

\subsection{The Promise of Agentic AI for Laboratory Automation}
The limitations of traditional DMTA implementation highlight the need for more intelligent approaches to laboratory automation and workflow coordination. While robotic systems have automated many routine tasks, they lack the contextual understanding and decision-making capabilities needed to optimize complex workflows like the DMTA cycle.

The emergence of agentic AI—artificial intelligence systems capable of autonomous reasoning, planning, and goal-directed behavior—offers transformative potential for pharmaceutical research\cite{hosseini2025role,boskabadi2025industrial,barra2025prompt,joshi2025comprehensive}. Unlike traditional AI applications that focus on pattern recognition or prediction, agentic AI systems can understand context, make decisions, and coordinate complex workflows without constant human supervision. This capability is particularly valuable in drug discovery, where researchers must integrate vast amounts of heterogeneous data, coordinate multiple experimental workflows, and make time-critical decisions that impact both scientific outcomes and business objectives.

Agentic AI represents a fundamental shift from passive AI tools to autonomous AI systems capable of goal-directed behavior, reasoning, and collaboration. Unlike traditional AI applications that respond to specific inputs with predetermined outputs, agentic AI systems can understand objectives, develop plans, adapt to changing circumstances, and coordinate with other agents to achieve complex goals.

In the context of scientific research, agentic AI systems possess several key capabilities that distinguish them from conventional automation. They can understand and interpret scientific literature, experimental protocols, and analytical data using natural language processing and domain-specific knowledge. They can reason about experimental design, predict likely outcomes, and suggest alternative approaches when initial strategies fail. They can communicate effectively with human researchers, other AI agents, and laboratory instrumentation using appropriate interfaces and protocols.

Most importantly, agentic AI systems can maintain persistent goals and adapt their strategies as new information becomes available. This capability is crucial for managing the iterative nature of scientific research, where hypotheses must be refined based on experimental results and unexpected findings often lead to new research directions.

\section{Tippy: A Multi-Agent System for Drug Discovery}
\label{sec:method}

\subsection{Agent Specialization and Domain Expertise}

\begin{figure}[H]
  \centering
  \includegraphics[width=1.0\textwidth]{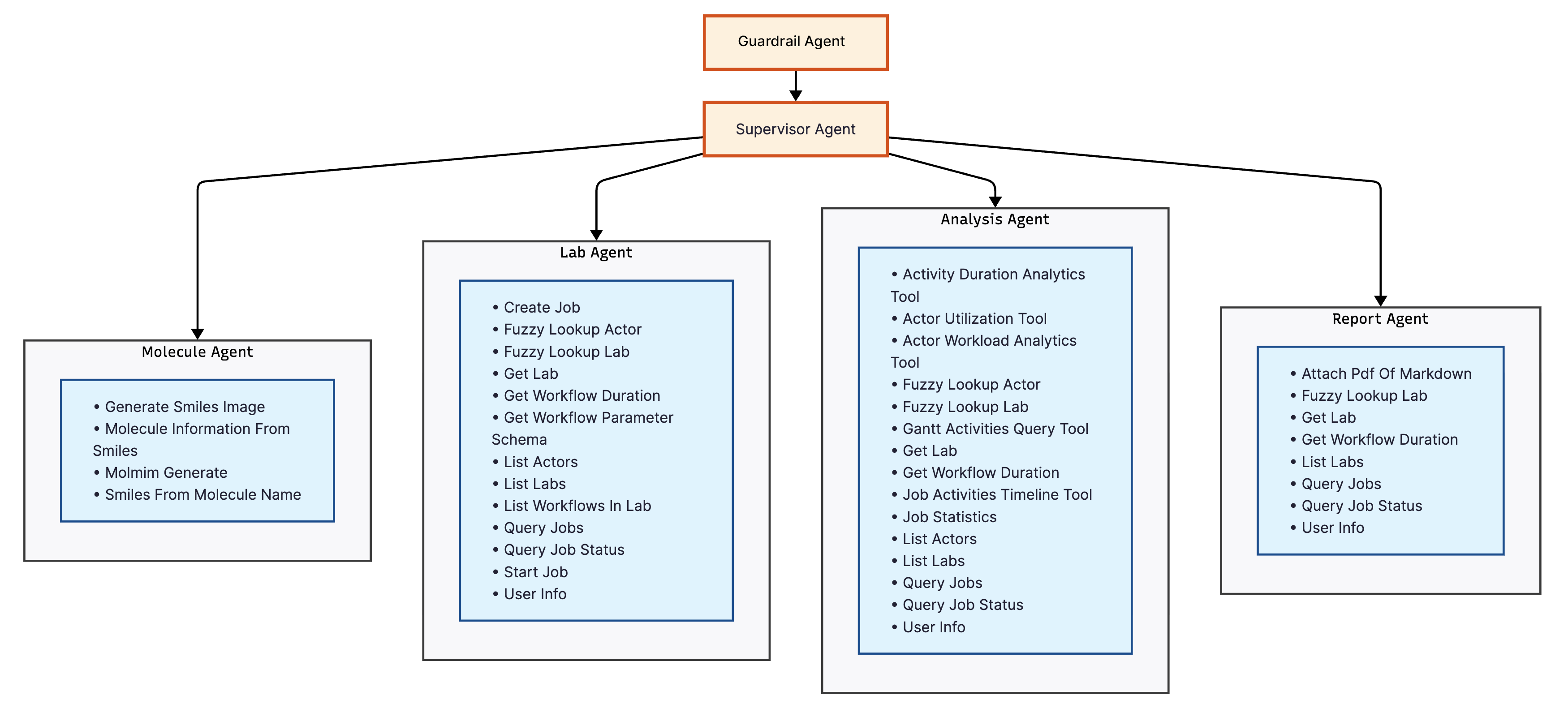} 
  \caption{Tippy's multi-agent architecture showing specialized agents (Supervisor, Molecule, Lab, Analysis, Report, and Safety Guardrail) with their associated 
  tools for laboratory automation and drug discovery workflows.}
  \label{fig:fig1}
\end{figure}

The complexity of modern pharmaceutical research requires specialized expertise across multiple domains, from computational chemistry to synthetic organic chemistry to analytical biochemistry. No single AI system can master all these domains to the depth required for autonomous operation. Instead, our approach leverages specialized agents, each optimized for specific aspects of the drug discovery process.

Figure~\ref{fig:fig1} illustrates the multi-agent architecture of Tippy, showing the five specialized agents and their associated tools. The Supervisor Agent acts as the central coordinator, orchestrating workflows between five domain-specific agents: the Molecule Agent for computational chemistry tasks, the Lab Agent for laboratory automation and instrumentation, the Analysis Agent for data processing and statistical analysis, and the Report Agent for documentation and communication. The Safety Guardrail Agent provides critical oversight across all operations. Each agent is equipped with specialized tools tailored to their domain expertise, enabling autonomous operation within their respective areas while maintaining seamless coordination through the supervisor

The Supervisor Agent serves as the central coordinator and primary interface for human researchers within the multi-agent system. It maintains overall project context, understands research objectives and timeline constraints, and orchestrates workflow execution across all specialized agents. The supervisor manages task delegation, determines which agent should handle specific requests based on task requirements, and facilitates seamless handoffs between agents as workflows progress through different DMTA phases. Additionally, it provides strategic guidance to human researchers, suggesting optimal next steps based on experimental results and project goals while maintaining visibility into resource allocation and system performance.

The Molecule Agent specializes in generating molecular structures and converting chemical descriptions into standardized formats. It can look up known molecules, suggest similar compounds, and optimize them for drug-likeness properties such as QED (quantitative estimate of drug-likeness) and logP (partition coefficient). The agent converts molecular names to SMILES notation and can generate molecular images when needed, serving as the primary driver of the Design phase in DMTA cycles.

The Lab Agent functions as the primary interface to the Artificial platform\cite{fehlis2025acceleratingdrugdiscoveryartificial}, managing HPLC analysis workflows, synthesis procedures, and laboratory job execution. It can create and start jobs, query job status and results, manage workflow parameters, and coordinate laboratory resources. The agent coordinates the Make and Test phases of DMTA cycles, ensuring efficient execution of synthesis and analytical workflows while maintaining quality and safety standards.

The Analysis Agent serves as a specialized data analyst, processing job performance data and extracting statistical insights from laboratory workflows. It has access to advanced analytics tools including activity duration analysis, actor workload monitoring, and job timeline tracking. It uses retention time data from HPLC analysis to guide molecular design decisions, as retention time correlates with key drug properties. It drives the Analyze phase of DMTA cycles, converting raw data into scientific understanding and decision-making guidance that feeds directly back into the Design phase.

The Report Agent acts as a documentation specialist, generating summary reports and detailed scientific documentation from experimental data. It can convert findings into PDF formats and attach reports directly to job results in the Artificial Lab suite, ensuring that insights from experiments are properly captured and shared with research teams.

The Safety Guardrail Agent operates as a critical safety oversight component, validating all user requests for potential safety violations before they are processed by other agents. This specialized agent analyzes requests for dangerous chemical reactions, unauthorized system access, procedures that could cause harm, and synthesis of controlled substances. It operates using a fast, lightweight model to provide immediate safety validation without impeding workflow efficiency, ensuring that all laboratory operations maintain the highest safety standards while allowing normal research activities to proceed unimpeded.

\subsection{Coordination and Collaboration Mechanisms}
Individual agent capabilities, while impressive, achieve their full potential only through effective coordination and collaboration. Our multi-agent framework implements several mechanisms to ensure seamless cooperation between specialized agents while maintaining clear responsibilities and avoiding conflicts.

A hierarchical coordination structure provides overall direction and conflict resolution through a Supervisor Agent that acts as the central coordinator and primary interface for human researchers. It understands project objectives, manages task delegation to specialized agents, and orchestrates handoffs between agents as workflows progress through different phases. The supervisor maintains overall context about ongoing projects and determines which specialized agent should handle specific requests based on the nature of the task. Additionally, the Supervisor Agent can guide human researchers on optimal next steps, suggesting strategic directions based on experimental results and project goals.

Dynamic handoff mechanisms allow agents to transfer control smoothly as workflows progress through different phases. When the Molecule Agent completes a design task, it can seamlessly transfer relevant information to the Lab Agent for synthesis planning. Similarly, when experimental data becomes available, the Lab Agent can hand off to the Analysis Agent for interpretation without human intervention.

Shared knowledge bases ensure that all agents have access to relevant information about ongoing projects, experimental results, and historical data. This shared context enables agents to make informed decisions and avoid duplicating work or pursuing contradictory objectives.

Collaborative decision-making protocols allow multiple agents to contribute to complex decisions that span multiple domains. For example, selecting the next compounds to synthesize might require input from the Molecule Agent (synthetic accessibility), Lab Agent (resource requirements), and Analysis Agent (information value based on existing data).

\subsection{Tippy Implementation on the Artificial Platform}

To demonstrate the practical application of agentic AI in pharmaceutical research, we have developed "Tippy," a comprehensive multi-agent system deployed on the Artificial platform. Tippy represents the first production-ready implementation of specialized AI agents designed specifically for automating the DMTA cycle, providing a concrete example of how agentic AI can transform laboratory workflows.

Tippy employs synthetic use cases that mimic early drug discovery workflows, recreating the complex decision-making processes that occur in real pharmaceutical laboratories. These controlled scenarios allow the AI agents to be tested and refined while demonstrating their effectiveness in coordinating DMTA cycles. The system uses retention time from HPLC analysis as a key scoring metric to guide molecule generation, creating a closed-loop learning system that mirrors the iterative refinement process used by human medicinal chemists.

\begin{figure}[H]
  \centering
  \includegraphics[width=1.0\textwidth]{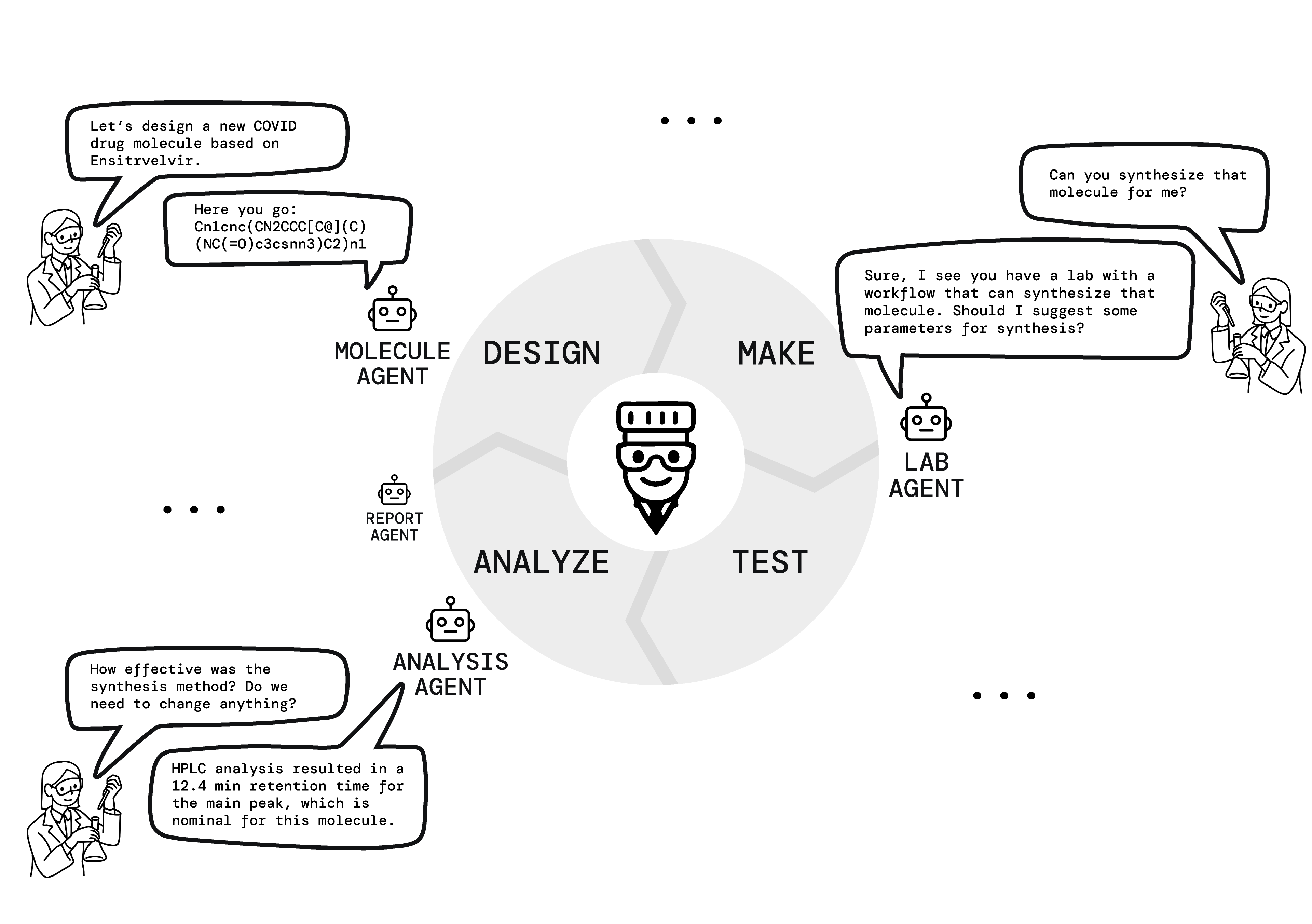} 
  \caption{It shows how Tippy's agents work with researchers in each step of the DMTA cycle. The Molecule Agent helps with design, the Lab Agent manages synthesis and 
  testing, the Analysis Agent processes results, and the Report Agent documents findings, creating a continuous feedback loop for drug discovery.}
  \label{fig:fig2}
\end{figure}

Figure~\ref{fig:fig2} illustrates the practical implementation of Tippy's agents within the DMTA cycle, showing how human researchers interact with specialized agents at each phase. The diagram demonstrates the seamless coordination between agents and users, from initial molecular design through synthesis, testing, and analysis, with continuous feedback loops that enable iterative optimization of drug candidates.

In a typical DMTA cycle within Tippy, the workflow unfolds with seamless agent coordination: The Molecule Agent initiates the Design phase by generating new molecular structures and optimizing them for drug-like properties. The Lab Agent then manages the Make phase, creating synthesis jobs and coordinating HPLC analysis workflows on the Artificial platform. During the Test phase, the Lab Agent orchestrates experimental execution while the Analysis Agent processes incoming data in real-time. Finally, in the Analyze phase, the Analysis Agent extracts insights from retention time data and other analytical results, feeding this information back to the Molecule Agent for the next design iteration. Throughout this process, the Report Agent documents the entire cycle, while the Supervisor Agent orchestrates handoffs and provides strategic guidance to both the specialized agents and human researchers.

\subsection{Integration with Laboratory Infrastructure}

Successful implementation of agentic AI in pharmaceutical research requires seamless integration with existing laboratory infrastructure, data systems, and workflows. Our approach emphasizes compatibility with current practices while enabling gradual transition to more automated and intelligent workflows.

All agent tools are implemented through the Model Control Protocol (MCP), which provides a standardized interface for AI agents to interact with laboratory systems, computational tools, and data resources. MCP enables consistent tool discovery, parameter validation, and error handling across all agent interactions, ensuring reliable and secure access to laboratory capabilities while maintaining clear boundaries between agent responsibilities and system resources.

The system integrates with laboratory information management systems (LIMS), electronic laboratory notebooks (ELN), and analytical instrument data systems through standardized APIs and data formats. This integration ensures that AI agents can access historical data, update records automatically, and coordinate with existing quality control and regulatory compliance systems.

Compatibility with common analytical instruments and automation platforms enables agents to control experiments directly when appropriate or provide detailed guidance to human operators when direct control is not feasible. The system can interpret instrument protocols, monitor experimental progress, and troubleshoot common issues without human intervention.

Integration with computational chemistry platforms and molecular databases allows agents to leverage existing tools and datasets while contributing new insights and predictions. Rather than replacing existing software, our agents serve as intelligent interfaces that can orchestrate complex computational workflows and interpret results in the context of broader research objectives.

\subsection{Impact of Tippy on Drug Discovery Workflows}
Tippy transforms pharmaceutical research by bridging communication barriers between computational chemists, synthetic chemists, and analytical scientists who traditionally work in isolation. The specialized agents provide seamless coordination across disciplines while enabling predictive research strategies that forecast experimental outcomes and adapt to unexpected results in real-time. Additionally, Tippy can guide users through complex workflows, suggesting optimal next steps based on experimental results and project goals. 

\section{Conclusions and Future Directions}

The implementation of Tippy represents a fundamental shift in drug discovery research, moving from manual coordination toward intelligent orchestration by autonomous AI agents. Our multi-agent framework demonstrates that specialized AI systems can automate routine tasks while reasoning about complex scientific problems and accelerating the DMTA cycle with maintained scientific rigor.

The success of agentic AI opens opportunities for enterprise-scale implementation, enabling coordination of multiple DMTA cycles and integration with business intelligence platforms. Future development must ensure AI agents enhance rather than replace human expertise, maintaining human responsibility for critical decisions while providing transparency for regulatory confidence.

The future of drug discovery will be characterized by human-AI collaboration, where autonomous agents handle workflow coordination and execution while human researchers provide strategic direction and ethical oversight. This partnership between human and artificial intelligence offers significant potential for developing therapeutics more efficiently while maintaining the quality and safety standards essential for patient care\cite{li2023call}.

%Bibliography
\bibliographystyle{unsrt}  
\bibliography{references}

\appendix
\section{Appendix: Real-World DMTA Cycle Demonstration with Tippy}

The following example demonstrates an actual interaction between a researcher and Tippy during a complete DMTA cycle, showcasing the capabilities of Artificial's laboratory automation platform for COVID drug discovery.

\subsection{Background Context}
The Design, Make, Test, Analyze (DMTA) cycle is a foundational process in pharmaceutical R\&D, powering the rapid iteration and refinement of drug candidates. By systematically generating new molecules, synthesizing them in the lab, testing their properties, and analyzing results, scientists are able to accelerate discovery timelines and improve success rates. This demonstration shows how Tippy, an AI assistant powered by specialized agents, can streamline each step of the DMTA cycle.

Artificial integrates Tippy as a helpful lab assistant, named in honor of pipette tips. With Tippy, scientists, operators and lab technicians can access the agentic power of specialized scientific agents and tools for laboratory automation and workflow coordination.

\subsection{DMTA Cycle Demonstration}

\textbf{Tippy:} "What would you like to do today?"

\textbf{User:} "I would like you to propose a new COVID drug molecule based on Ensitrelvir."

\textbf{Tippy:} "Wonderful. Here are some molecules generated by MolMIM:

\begin{itemize}
\item \texttt{CCN(CC(C)(C)C)c1ccc2c(=O)n(C)c(=O)n(C)c2n1}
\item \texttt{CCN(CC(C)(C)C)c1nc2c(c(=O)n(C)c(=O)n2C)n1C}
\item \texttt{CN(CCC(C)(C)C)c1nc2c(c(=O)n(C)c(=O)n2C)n1C}
\end{itemize}

Would you like to visualize their structures or synthesize the molecules?"

\textbf{User:} "Yes. Please show me the first molecule's structure."

\textbf{Tippy:} "Here it is:"

\begin{figure}[ht]
  \includegraphics[width=0.2\textwidth]{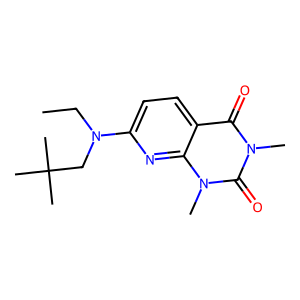} 
\end{figure}

Here, a scientist is interacting with Tippy to generate novel molecules derived from a known COVID treatment and then leveraging tools to automatically generate a visual representation of the new molecule. The next interactions demonstrate Tippy's automated synthesis experiment parameter optimization.

\textbf{Tippy:} "Would you like to synthesize one or more of these molecules?"

\textbf{User:} "Yes, synthesize the first molecule."

\textbf{Tippy:} "I see you have a Synthesis Lab with a Synthesis Workflow in it. To synthesize the first molecule, here are the parameters I recommend:

\textit{Note: Specific synthesis parameters have been omitted from this publication due to confidentiality. The actual demonstration included detailed specifications for target molecule, building blocks, reagents, solvents, temperatures, reaction times, and purification conditions - all automatically optimized by Tippy based on the molecular structure and laboratory capabilities.}

Do you want to change any of these or would you like to proceed with scheduling this work?"

\textbf{User:} "Yes. Thank you, please schedule it to run ASAP."

\textbf{Tippy:} "Great. I have scheduled and started the Synthesis Workflow with the parameters we discussed. The scheduler is predicting that the workflow will take 2 hours."

Tippy, in coordination with Artificial's LabOps agents, has automatically created, configured, and optimally scheduled the synthesis workflow to run in an available lab. The synthesis workflow is now ready to execute and produce the molecule.

\textit{Synthesis workflow completes}

\textbf{User:} "I have synthesized the molecule and I would like to check my results for yield and purity."

\textbf{Tippy:} "Great. I will pass the sample molecule to the Analysis Lab, and run the HPLC workflow to check the purity of your synthesized compounds by analyzing the retention time. Given your molecule, I recommend these parameters:

\textit{Note: Specific HPLC parameters have been omitted from this publication due to confidentiality. The actual demonstration included detailed analytical conditions such as column specifications, mobile phase compositions, gradient programs, flow rates, detection wavelengths, sample preparation protocols, and quality control measures - all automatically optimized by Tippy for the target compound.}

Would you like to change any of these or should I proceed with scheduling a job in the Analysis Lab on an HPLC?"

Tippy and the Artificial agents have automatically configured an HPLC run to measure yield and purity of the molecule, demonstrating the Test phase of the DMTA cycle.

\textbf{User:} "Run it."

\textbf{Tippy:} "I have scheduled and started to run the HPLC workflow with your samples and the parameters I suggested. The HPLC workflow will take 45 minutes to complete."

\textit{HPLC analysis completes}

\textbf{User:} "How's that job doing?"

\textbf{Tippy:} "The HPLC has completed and the results are within the expected range of synthesis efficiency. The HPLC chromatogram showed a primary peak at 8.5 minutes with a purity of 95.3\%. The synthesis yielded 48 mg of material with 72\% yield. No major side products observed. I have uploaded the results to cloud storage for further downstream analytics. Would you also like me to add a summary of the results into the job's associated data store?"

\textbf{User:} "Yes please."

\textbf{Tippy:} "If you'd like to improve the synthesis we could retry with different parameters."

\textbf{User:} "No thanks, this looks great!"

\subsection{Summary}
This demonstration showcases Artificial's lab orchestration platform empowering pharmaceutical organizations to execute the full DMTA cycle with unprecedented speed, precision, and intelligence. From AI-assisted molecule design to automated synthesis planning, workflow scheduling, experimental execution, and results analysis---each phase is streamlined through the collaboration of intelligent agents like Tippy. The seamless coordination between lab operations and cloud-based data infrastructure ensures that every insight is captured, stored, and ready for downstream analysis, enabling scientists to make faster, better-informed decisions.

\end{document}